\documentclass[conference]{IEEEtran}
\IEEEoverridecommandlockouts
\usepackage[numbers]{natbib}
\usepackage{amsmath,amssymb,amsfonts}
\usepackage{algorithmic}
\usepackage{graphicx}
\usepackage{textcomp}
\usepackage{xcolor}
\usepackage{tcolorbox}
\usepackage{paralist}

\def\BibTeX{{\rm B\kern-.05em{\sc i\kern-.025em b}\kern-.08em
    T\kern-.1667em\lower.7ex\hbox{E}\kern-.125emX}}
\begin{document}

\title{Towards Safety-Aware Mutation Testing for Autonomous Driving Systems
}

\author{\IEEEauthorblockN{Donghwan Shin}
\IEEEauthorblockA{\textit{The University of Sheffield} \\
ORCID: 0000-0002-0840-6449 \\
}
}

\maketitle

\begin{abstract}
Simulation-based testing is essential for ensuring the safety of Autonomous Driving Systems (ADS), yet the community lacks a systematic criterion for determining when we can safely stop additional test scenario generation. Existing coverage metrics typically focus on individual component reliability or treat the ADS as a black box, failing to capture certain component interactions that cause most ADS accidents. While traditional mutation testing provides a falsifiable measure of test adequacy, directly porting code- and deep learning model-level mutations to the corresponding modules of ADS is insufficient.

In this vision paper, we propose a paradigm shift toward Safety-Aware Mutation Testing (SAMT). Unlike traditional mutation testing, which creates mutants (i.e., faulty versions of the software under test) by injecting artificial faults into individual components, SAMT systematically injects temporally bounded faults into the messages exchanged between ADS modules to simulate realistic interaction failures. To ensure these mutants represent genuine hazards, we propose deriving mutant generation rules directly from top-down safety engineering frameworks, such as System-Theoretic Process Analysis (STPA). By embedding systems thinking into the mutation testing pipeline, SAMT provides a rigorous mechanism for evaluating test adequacy, enabling automated scenario generation, and guiding ADS repair. We also outline critical open challenges.
\end{abstract}


\section{Introduction}

Autonomous Driving Systems (ADS) are safety-critical software systems, and assuring their safety requires rigorous testing. Simulation-based testing has been widely studied as a mechanism for automatically generating diverse, potentially critical driving scenarios \cite{9763411,zhong_survey_2021}. However, despite significant progress in scenario generation, the community still lacks a systematic, objective criterion for determining when test generation can be safely stopped.

What makes it more challenging is that safety is an emergent system-level property. It depends on the interactions among multiple modules of ADS (e.g., perception, prediction, planning, and control modules), not merely on the standalone reliability of each component. As a result, test adequacy criteria that focus exclusively on individual modules would fail to capture system-level safety risks arising only from component interactions.

Mutation testing \cite{jia_analysis_2011,papadakis_mutation_2019} offers an appealing foundation for adequacy assessment because it makes test effectiveness claims falsifiable \cite{papadakis_mutation_2019}. By systematically injecting small faults into a software system to create mutated versions (mutants) and checking whether a test suite distinguishes the mutants from the original system, mutation testing provides concrete evidence of which faults the tests can and cannot detect. A test suite that fails to distinguish specific mutants reveals precise inadequacies and, by implication, potential blind spots with respect to corresponding real faults.

However, directly porting mutation testing for source code \cite{pit,mutest-rs} or Deep Learning (DL) models \cite{hu_deepmutation_2019,humbatova_deepcrime_2021} to ADS is insufficient. This is because ADS have a mix of traditional software and DL components, and their interactions during continuous closed-loop execution, rather than the individual components, can lead to safety violations. In this vision paper, we argue for a paradigm shift from component-level code/DL mutation to system-level, safety-aware mutation. Our vision statement is as follows:

\begin{tcolorbox}
By embedding safety engineering principles into the mutation testing pipeline, we can systematically evaluate test adequacy and guide test scenario generation, shifting the ADS testing focus from individual component reliability to system safety.
\end{tcolorbox}

The rest of the paper is organised as follows. Section~\ref{sec:background} provides background on mutation testing, ADS architecture, existing coverage metrics, and safety engineering principles. Section~\ref{sec:vision} outlines our vision for Safety-Aware Mutation Testing (SAMT) for ADS, detailing the key innovations and process. Section~\ref{sec:challenges} discusses open challenges and future research directions for realising this vision. Finally, Section~\ref{sec:conclusion} concludes the paper.

\section{Background}
\label{sec:background}

\subsection{Mutation Testing}
\label{sec:mutation_background}
Mutation testing is a fault-based technique designed to evaluate the adequacy of a test suite by systematically injecting artificial defects, called \textit{mutants}, into a software artefact using predefined transformation rules known as \textit{mutation operators} (e.g., replacing `+' with `-' or deleting a program statement). 
A test suite is then executed against both the original program and its mutants; if a test case yields a different observable outcome for a mutant compared to the original, that mutant is considered \textit{killed}. 
Killing all mutants provides a falsifiable assertion of the test suite's effectiveness at detecting real-world faults \cite{papadakis_mutation_2019}. 
A key hypothesis underpinning this technique is the \textit{coupling effect}, which posits that tests capable of killing simple mutants are also likely to detect more complex, real faults \cite{coupling_effect}. 
This effect has been empirically validated in traditional software testing contexts \cite{just_are_2014,gopinath_mutations_2014,papadakis_are_2018}. 

With the advances of DL, researchers have developed specialised mutation techniques for DL models (e.g., perturb neural network weights, remove neurons, or manipulate training data) \cite{hu_deepmutation_2019,humbatova_deepcrime_2021}. 
While these techniques could effectively evaluate the localised robustness of an isolated DL model, they are fundamentally inadequate for testing an entire ADS. 

A persistent challenge in the mutation testing paradigm, regardless of the mutated artefact, is the \textit{equivalent mutant problem}: mutants that are syntactically modified but remain semantically identical to the original. The presence of equivalent mutants can lead to false negatives in test adequacy assessments. However, determining whether a mutant is equivalent is generally undecidable \cite{budd1982two}. 
In practice, researchers have developed various heuristics to identify and filter out equivalent mutants \cite{hierons_using_1999,papadakis_trivial_2015,tian_large_2024}. 

\citet{10132190} recently proposed Property-Based Mutation Testing (PBMT), which relates mutation to system-level safety requirements. Targeting Simulink data-flow models, PBMT effectively filters out semantically irrelevant faults by anchoring kill criteria to explicit temporal specifications. However, its scope remains confined to internal block or signal perturbations within monolithic architectures, overlooking emergent hazards arising from degraded inter-module communication in modern modular systems. Furthermore, PBMT does not systematically derive fault injections from formal hazard analyses, nor does it account for closed-loop control dynamics. These limitations underscore a critical need for mutation testing frameworks that operate at the system-interaction level and are grounded in architectural safety constraints rather than in isolated component outputs.

\subsection{ADS Architecture}
Broadly speaking, ADS can be categorised into End-to-End (E2E) and module-based. E2E ADS, such as NVIDIA Dave2 \cite{dave2}, treat the system as a single DL model that processes sensor data (e.g., camera images) and directly outputs vehicle control commands (e.g., steering and acceleration). Despite its simplicity, it requires substantial training data to achieve acceptable performance and suffers from limited explainability. 

In contrast, module-based ADS, such as Apollo \cite{fan2018baidu} and Autoware \cite{kato2018autoware}, divide the driving task across multiple dedicated modules (e.g., perception, prediction, planning, and control). This modularity offers essential explainability and debugging capabilities for safety-critical systems. 

Communication among these modules is implemented via a publish-subscribe architecture. Each module publishes its output as messages to specific topics and subscribes to others to receive the necessary messages. For example, the perception module publishes bounding boxes for detected objects, which the planning module subscribes to in order to compute safe trajectories. This design enables non-blocking communication and module hot-swapping, which are vital for the real-time operation of an ADS \cite{macenski_robot_2022}.

Considering the industrial relevance of module-based ADS, we focus on them in this paper. Nevertheless, the proposed vision can be extended to E2E ADS as long as the notion of messages between functionally distinct components can be defined.

\subsection{Coverage Metrics for ADS Testing}
Although traditional code coverage metrics (e.g., statement or branch coverage) can assess the adequacy of test suites for deterministic software components in an ADS, they are insufficient for DL-enabled components whose logic is derived from training data. 

To address this, researchers have proposed structural coverage metrics for DL models, such as neuron coverage \cite{deeptest} and its variants \cite{10.1145/3358233}. However, these metrics still focus on isolated components and fail to account for module interactions.

More recently, researchers have proposed ADS-specific coverage metrics, such as scene coverage \cite{woodlief_s3c_2024} and scenario coverage \cite{schallau_stars_2024}. However, these approaches largely treat the ADS as a black box. Because they do not account for the ADS under test's internal architectural state or its specific fault-handling mechanisms, they cannot verify that the system's internal safety boundaries have been adequately exercised.

\subsection{Safety Engineering and Systems Thinking}
\label{sec:safety_engineering}
In complex cyber-physical systems, safety and reliability are distinct properties \cite{leveson_engineering_2011}. A component can be highly reliable (meeting its specifications) but still lead to an unsafe system state. For example, consider a perception module that incorrectly identifies a pedestrian as a shadow due to glare. If the planning module's logic strictly follows the ``drive through shadows'' policy, the vehicle will hit the pedestrian. From a reliability standpoint, one might argue the perception module ``failed'' and requires more training data. However, perception can never be 100\% perfect in all edge cases in the real world.

To address this issue, modern safety engineering calls for \textit{systems thinking}, viewing safety as a control problem rather than a component reliability problem. This involves defining \textit{safety requirements} (or \textit{safety constraints}), i.e., rules that specify what the system must not do (e.g., the vehicle must never proceed if the path's safety is uncertain). Frameworks like System-Theoretic Process Analysis (STPA) \cite{leveson_engineering_2011} are used to systematically identify Unsafe Control Actions (UCAs), instances where a module's output can lead to the violation of safety constraints (e.g., an instance where the planning module issues a ``maintain speed'' command when the perception module's object classification confidence falls below a safe threshold).
By applying systems thinking, we can shift focus from fixing the perception module to designing safer component interactions, such as enforcing high-confidence uncertainty propagation to the planner to safely handle perception limitations.

\section{Vision: Safety-Aware Mutation Testing}
\label{sec:vision}

\begin{figure}
    \centering
    \includegraphics[width=1\linewidth]{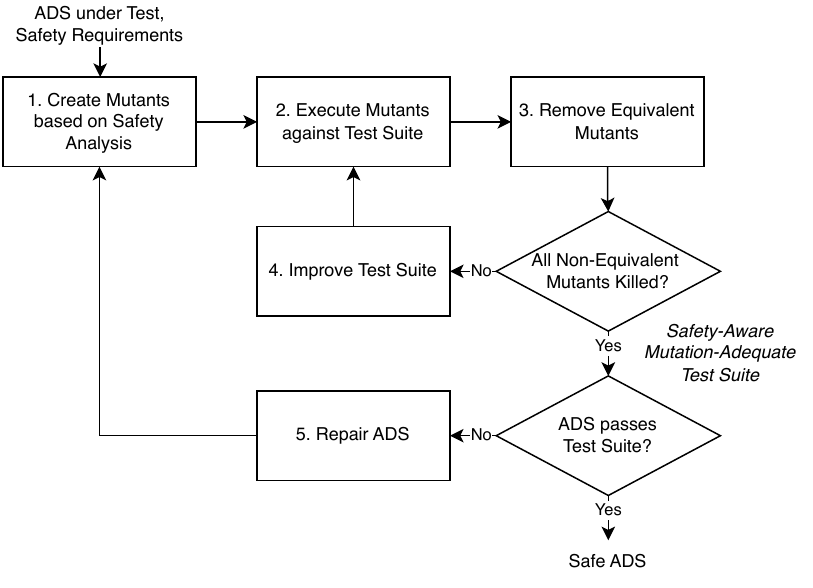}
    \caption{Safety-Aware Mutation Testing Process Overview}
    \label{fig:overview}
\end{figure}

This section outlines the vision for Safety-Aware Mutation Testing (SAMT) for ADS. The overall process, illustrated in \figurename~\ref{fig:overview}, consists of the following steps.
\begin{enumerate}[(1)]
    \item \textbf{Mutant Generation}: Apply a library of safety-aware mutation operators (\S~\ref{sec:mutation_operators}) to systematically create mutants by injecting faults into the messages shared between ADS modules, simulating realistic interaction faults (\S~\ref{sec:message_mutation}). 
    \item \textbf{High-Fidelity Execution}: Execute the mutants against the test suite in a high-fidelity simulation environment (e.g., CARLA \cite{dosovitskiy_carla_2017}). To reduce computational overhead, execution can be stopped as soon as a mutant is definitively killed.
    \item \textbf{Equivalent Mutant Removal}: Analyse the execution results to determine whether mutants are killed, not killed, or equivalent (\S~\ref{sec:eq_mutant_detection}). 
    \item \textbf{Test Suite Improvement}: If all non-equivalent mutants are killed, the test suite is deemed adequate. Otherwise, surviving mutants provide concrete evidence of test suite blind spots, guiding further scenario generation (\S~\ref{sec:test_improvement}).
    \item \textbf{Safety Assessment \& Repair}: Once the test suite is adequate, verify the original ADS against it. Passing implies the ADS satisfies the defined safety requirements, while failing highlights violations that guide downstream debugging and architectural repair (\S~\ref{sec:applications}).
\end{enumerate}

In the following subsections, we elaborate on the above steps, along with key innovations.

\subsection{Shift from Components to Their Messages}
\label{sec:message_mutation}
A key innovation in our vision is the shift from mutating each component (e.g., the planning module's source code) to mutating the \textit{messages} exchanged between ADS modules. For example, we can mutate the perception module's output by removing detected objects, adding phantom objects, or introducing spatial perturbations to bounding boxes. This directly simulates the interaction faults and degraded feedback loops that cause system-level accidents.

Furthermore, given the closed-loop, continuous-control nature of ADS, we can extend mutation operators with \textit{temporal properties}, such as the duration and triggering condition of their effects. For example, a mutation operator could suppress an obstacle specifically between simulation timestamps $t=2$ and $t=3$. This allows SAMT to additionally simulate subtle functionality limitations that are critical for realistic ADS evaluation.

\subsection{Mutation Operators Derived from Safety Analysis}
\label{sec:mutation_operators}
Mutation operators must represent genuine safety threats rather than arbitrary data perturbations. We propose systematically deriving them through top-down safety engineering. 

First, hazard analysis techniques like STPA are applied to system-level safety constraints to identify abstract Unsafe Control Actions (UCAs) and their underlying causal factors, such as \textit{``missing vehicle-in-front data from the perception module''}. 
These causal factors can then be directly mapped to concrete, parameterised message-level mutation operators. For the factor above, the corresponding mutation operator intercepts the perception-to-planning topic and removes a parameterised entity. By grounding mutation operators in formal hazard analysis, SAMT ensures every generated mutant simulates a realistic component interaction failure.

\subsection{Equivalent Mutant Detection via Propagation Analysis}
\label{sec:eq_mutant_detection}
The equivalent mutant problem (\S~\ref{sec:mutation_background}) can be exacerbated by the complex safety mechanisms within ADS, which can often absorb internal faults without causing safety violations. To resolve this problem, we propose evaluating mutants using \textit{propagation analysis} of the continuous control states:

\begin{itemize}
\item \textbf{Weakly killed:} The mutated message successfully propagates through and alters the control module's output state (e.g., the steering angle or throttle trace diverges from the original ADS execution), but results in no difference in safety outcomes compared to the original ADS.
\item \textbf{Strongly killed:} The state divergence forces the mutant to violate a safety constraint that was otherwise maintained in the original execution.
\end{itemize}

If a mutant is never weakly killed by any test scenario in the suite (i.e., the control traces remain identical to the original executions), we can approximate that the mutant is equivalent (safe-by-design). Otherwise, the mutant is killable, meaning a safety-adequate test suite must contain at least one scenario that strongly kills it.

\subsection{Mutation-Guided Test Suite Improvement}
\label{sec:test_improvement}
If a non-equivalent mutant survives, it provides concrete evidence of a specific inadequacy in the test suite. 
We can automate the generation of new test scenarios that specifically target the surviving mutant, thereby iteratively improving adequacy. 
Search-Based Software Testing (SBST) \cite{mcminn_search-based_2004} can be effective here. 
SBST has been successfully used in traditional mutation testing to automatically generate unit tests that maximize the likelihood of killing target mutants \cite{papadakis_mutation_2012,fraser_achieving_2015}. 
Applying this to an ADS may require adapting the fitness function to search the continuous ODD parameter space (e.g., weather, NPC trajectories) to efficiently find the environmental entities that escalate a weak kill into a strong kill.

\section{Open Challenges}
\label{sec:challenges}

This section outlines several open challenges and a future research agenda for realising the vision.

\subsection{Empirical Validation of the Coupling Effect in ADS}
The coupling effect is a fundamental hypothesis in mutation testing, but it has primarily been validated only in traditional software contexts. In the ADS context, where safety violations often arise from complex component interactions rather than isolated failures, it is crucial to empirically validate whether this effect still holds. Future research must design controlled experiments to assess whether tests that kill simple, single-message mutants are reliably capable of detecting more complex, real-world interaction faults. While recent work \cite{chen_comprehensive_2025} has presented initial empirical studies of real-world bug-fix patterns in ADS, the community urgently needs a comprehensive, open-source catalogue of real-world ADS faults, similar to Defects4J for Java \cite{just_defects4j_2014}. This foundational validation is essential to justify the use of message-level mutation as a definitive adequacy criterion for ADS certification.

\subsection{Comprehensive Mutation Operator Library for ADS}
\label{sec:op_libs}
While specific safety requirements might vary between ADS architectures, the underlying physics and closed-loop control dynamics share commonalities. A major future direction is for the community to establish a standardised set of generic ADS safety requirements, which can serve as the foundation for a universally applicable library of parameterised mutation operators (e.g., spatial offset errors, temporal delays, classification degradation).

To achieve this, we must systematically link abstract safety analysis artefacts to executable code. This requires developing, for example, an \textit{STPA-to-Mutant Mapping Layer}, where abstract UCAs are translated into formal, parameterised interfaces that can be automatically instantiated into concrete safety-aware mutants.

\subsection{Efficient Equivalent Mutant Detection}
\label{sec:efficient_eq_detection}
Detecting ``safe-by-design'' equivalent mutants via full-system propagation analysis requires executing the high-fidelity simulation, which is computationally expensive. A critical open challenge is developing advanced, lightweight propagation analysis techniques that can be applied before running all mutants against a test suite. 

In traditional software programs, for example, program slicing can be used to determine whether a mutated statement has the potential to affect output variables \cite{hierons_using_1999}. Similarly, causal data-flow analysis could identify ``masking'' effects without running a full simulation for some mutants. For example, if the ADS under test uses sensor fusion heavily, weighting LIDAR over camera inputs, mutants relevant to camera inputs would be masked.

\subsection{Downstream Applications: Fault Localisation and Repair}
\label{sec:applications}
Moving beyond test adequacy, the execution traces of safety-aware mutants offer rich data for downstream applications like automated fault localisation and system repair. In traditional software and DL models, mutation-based fault localisation \cite{papadakis_metallaxis-fl_2015,7372034,ghanbari_mutation-based_2023} identifies suspicious program statements and model layers by correlating them with killed mutants. While some ADS-specific fault localisation \cite{feng_rocas_2024,sun_acav_2024} and temporary repair techniques \cite{sun_fixdrive_2025} exist, they do not yet leverage mutation analysis. Future research should integrate SAMT traces (e.g., contrasting weakly killed vs strongly killed paths) with existing localisation approaches to pinpoint exactly which module interactions are failing in the original ADS. Furthermore, surviving mutants could serve as continuous fitness functions for automated program repair tools aimed at synthesising permanent, safe control-loop patches.

\subsection{Nondeterminism of ADS Simulation}
Although high-fidelity simulators like CARLA provide realistic environments, they exhibit inherent nondeterministic behaviour due to various reasons, including floating-point arithmetic errors, runtime scheduling variations, and game engine setup \cite{chance_determinism_2022,osikowicz_empirically_2025}. This nondeterminism can cause natural-state divergences, making it difficult to determine whether a mutant is killed. Future research must investigate statistical methods, such as threshold-based equivalence checking, to reliably assess propagation analysis over multiple simulator iterations with statistically rigorous confidence intervals.

\section{Conclusion}
\label{sec:conclusion}

The current landscape of ADS testing suffers from a common misunderstanding: the conflation of component reliability with system-level safety. As long as test adequacy criteria remain focused on isolated code structures or deep learning models, the community will struggle to objectively determine when scenario generation can be safely stopped. 

To bridge this gap, we propose a paradigm shift toward Safety-Aware Mutation Testing (SAMT). Instead of mutating static code or DL models, SAMT injects safety-critical faults into the messages communicated between ADS modules. This message-level approach captures a wide range of realistic problems. It naturally covers traditional component crashes (since a failed component simply stops sending messages or sends corrupted data), while also exposing the subtle interaction failures that traditional metrics miss. 

Realising this vision requires a collective community effort to overcome various open challenges, from empirically validating the coupling effect in ADS to managing simulation nondeterminism. Ultimately, embedding systems thinking into the mutation testing pipeline offers a falsifiable, objective foundation not only for assessing test adequacy but for driving the automated generation, localisation, and repair of next-generation autonomous systems. 

\section*{Acknowledgment}
This work was supported by the Institute of Information \& Communications Technology Planning \& Evaluation(IITP) grant funded by the Korea government(MSIT) (No. RS-2025-02218761, 50\%) and by the Engineering and Physical Sciences Research Council (EPSRC) [EP/Y014219/1]. 

\bibliographystyle{IEEEtranN}

\end{document}